\documentclass[%
 reprint,
 amsmath,amssymb,
 aps,
]{revtex4-2}

\usepackage{multirow}   % preamble에 추가\
\usepackage{graphicx}% Include figure files
\usepackage{dcolumn}% Align table columns on decimal point
\usepackage{bm}% bold math
\usepackage{tabularx}
\usepackage{booktabs}
\usepackage{float}
\makeatletter
\let\newfloat\newfloat@ltx
\makeatother
\usepackage{caption}
\usepackage{ragged2e}
\usepackage{hyperref}

\DeclareCaptionFormat{myformat}{\justifying #1#2#3}
\usepackage{xcolor,colortbl}
\usepackage[table]{xcolor}
\usepackage{algorithm}
\usepackage[english]{babel}
\usepackage{algpseudocode}
\usepackage{adjustbox}
\usepackage{makecell}
\usepackage{amsmath}

\usepackage{nicematrix}
\usepackage{adjustbox}
\usepackage{multirow}
\usepackage{makecell}
\usepackage[table]{xcolor}
\newcolumntype{Y}{>{\centering\arraybackslash}X}

\begin{document}

\preprint{APS/123-QED}

\title{Reducing Postselection Overhead in Magic-State Cultivation by In-Patch Multiplexing}% Force line breaks with \\
% \title{In-Patch Multiplexing for Magic-State Cultivation}

\author{Dongmin Kim, Jeonggeun Seo, and Youngsun Han}
 \thanks{youngsun@pknu.ac.kr}%
\affiliation{%
Department of AI Convergence, Pukyong National University, Busan 48513, South Korea
}%
\author{Aniket Patra}
\affiliation{%
Department of Physics, Indian Institute of Technology Kharagpur, Kharagpur 721302, West Bengal, India
}%

\date{\today}% It is always \today, today,

\begin{abstract}

Fault-tolerant quantum computing requires high-fidelity logical magic states for implementing non-Clifford operations. Magic-state cultivation provides a lower-overhead route to logical magic-state preparation, but its efficiency is limited by postselection loss during the early injection-and-cultivation stages. In this work, we propose an in-patch multiplexing scheme that uses early-stage idle resources within a single logical patch to create multiple local cultivation opportunities. A candidate that passes the early stages is forwarded to the standard escape pathway, while the escape stage and the decoder-based acceptance procedure are kept identical to those of the single-site baseline. Under a uniform depolarizing noise model with idle noise, the proposed protocol substantially reduces the injection-and-cultivation discard rate and the expected number of attempts required to obtain an accepted early-stage candidate. At a physical error rate of \(p=2\times10^{-3}\), the injection-and-cultivation expected attempts are reduced by \(45.46\%\) for \(d_1=3\) and by \(72.91\%\) for \(d_1=5\), relative to the single-site magic state cultivation baseline. In the direct full-cycle evaluation including escape, the expected attempts per kept logical output are further reduced by \(49.04\%\) for \(d_1=3\) and by \(78.69\%\) for \(d_1=5\) at the same physical error rate. The full-cycle cost curves are shifted toward smaller expected attempts, while the final logical-error behavior remains governed by the escape-stage gap threshold. These results show that in-patch multiplexing can reduce postselection overhead while preserving the standard magic-state cultivation framework.
\end{abstract}

\maketitle

%\tableofcontents

\section{Introduction}
Fault-tolerant quantum computing (FTQC) is a framework for reliable universal quantum computation in the presence of physical noise, based on encoding quantum information into error-correcting codes and processing it through fault-tolerant logical operations~\cite{PhysRevA.54.1098, PhysRevA.52.R2493, gottesman_theory_1998,preskill_reliable_1998, terhal_quantum_2015}.
Its theoretical foundation is the threshold theorem, which establishes that logical errors can be suppressed even as the computation scales, provided that the physical error rate remains below a finite threshold~\cite{knill_resilient_1998,aharonov_fault-tolerant_2008, 10.5555/2011665.2011666, PhysRevLett.98.190504}.
To achieve universality, however, fault-tolerant implementations must go beyond Clifford operations alone.
Because Clifford operations are not universal, and because a universal transversal encoded gate set is incompatible with a single quantum error-correcting code, FTQC must incorporate fault-tolerant non-Clifford resources~\cite{bravyi_universal_2005, eastin_restrictions_2009, zeng_transversality_2011}.

A standard approach to fault-tolerant non-Clifford computation is to prepare magic states and consume them during the computation~\cite{gottesman_demonstrating_1999, knill_quantum_2005, 10.5555/2012098.2012105, 10.1007/s11128-005-7654-8}. Bravyi and Kitaev showed that universal quantum computation can be achieved using Clifford operations, computational-basis state preparation and measurement, supplemented by suitable nonstabilizer ancilla states, thereby formalizing such resource states as magic states~\cite{bravyi_universal_2005}. A representative example is the T-type magic state used to implement the T gate, whose one-qubit form is 
\[
|T\rangle = T|+\rangle = \frac{|0\rangle + e^{i\pi/4}|1\rangle}{\sqrt{2}}
.\]
In fault-tolerant architectures, non-Clifford gates are commonly realized not through the same native fault-tolerant procedures used for Clifford operations, but by consuming pre-prepared T-type magic states via state injection or gate teleportation~\cite{gottesman_demonstrating_1999, zhou_methodology_2000}.

The central practical challenge is that the relevant resource is not merely a physical ancilla, but a high-fidelity logical magic state that can be consumed at the algorithmic level. Such states are not typically available directly from native noisy operations and must instead be produced through additional preparation protocols acting on lower-fidelity resource states. These protocols require extra encoded qubits, repeated stabilizer measurement or verification rounds, and dedicated factory subroutines, so the cost of magic-state production can account for a substantial fraction of the total spacetime overhead in leading fault-tolerant architectures~\cite{ogorman_quantum_2017, campbell_roads_2017, bravyi_magic-state_2012, chamberland_very_2020}. Improving the efficiency of logical magic-state preparation is therefore widely regarded as a central requirement for scalable fault-tolerant quantum computation.

A longstanding approach to this problem is magic-state distillation~\cite{bravyi_universal_2005, bravyi_magic-state_2012, campbell_bound_2010, campbell_magic-state_2012}. In distillation protocols, multiple noisy logical magic states are consumed in order to produce a smaller number of higher-fidelity output states using fault-tolerant Clifford operations and measurements. This framework has led to a broad class of protocols, including low-overhead constructions by Bravyi and Haah and multilevel distillation schemes by Jones, which significantly improved the resource scaling of high-fidelity magic-state preparation~\cite{bravyi_magic-state_2012, jones_multilevel_2013}. Even so, distillation remains resource-intensive because it fundamentally relies on multiple encoded input states and additional logical circuitry. 

\begin{figure}[!t]
    \centering{\includegraphics[width=0.455\textwidth]{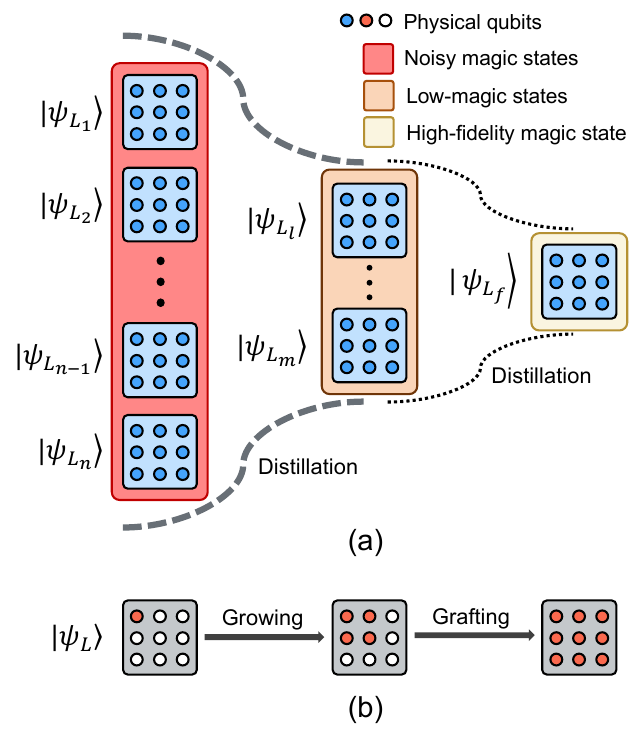}}
    \caption{
    Conceptual comparison between magic-state distillation and magic-state cultivation. (a) In magic-state distillation, multiple noisy logical magic states are consumed through one or more purification levels to produce a smaller number of higher-fidelity output states. (b) In magic-state cultivation, a single local magic state is grown through staged operations, such as injection, cultivation, and escape, to obtain a high-fidelity logical magic state within a larger logical patch.
    } \label{fig:DistillationAndCultivation}
\end{figure}

Recently, magic-state cultivation has been proposed as a lower-cost approach to logical magic-state preparation~\cite{gidney_magic_2024, rosenfeld2025magicstatecultivationsuperconducting, 9kys-3whh, p8tw-6kq9, hirano_efficient_2025}. Rather than consuming multiple encoded magic states in a factory-style distillation protocol, cultivation grows a single logical \(|T\rangle\) state within a single logical patch through a staged sequence of injection, cultivation, and escape. Gidney \textit{et al.} argued that this approach can substantially reduce the qubit-round cost required to reach intermediate logical error rates, suggesting a more direct route to practical high-fidelity \(T\)-state preparation~\cite{gidney_magic_2024}. The distinction between these two approaches is illustrated in Fig.~\ref{fig:DistillationAndCultivation}. At the same time, the cultivation protocol remains sequential and postselection-based, so its end-to-end efficiency depends not only on the cost of a successful trajectory but also on how often early-stage attempts are discarded and restarted.

\begin{figure}[!t]
    \centering{\includegraphics[width=0.48\textwidth]{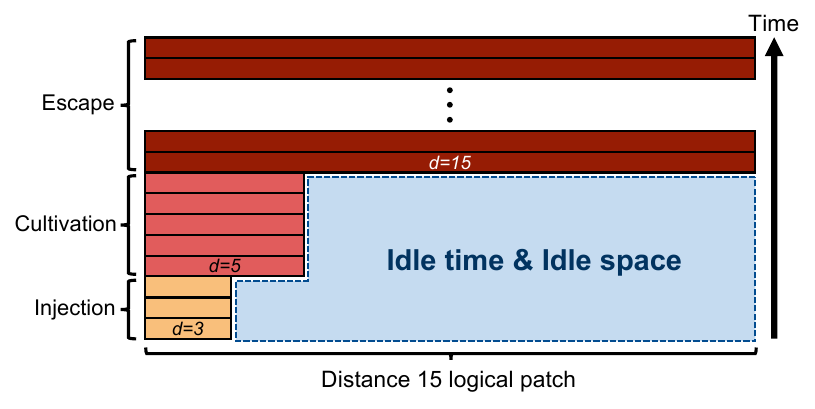}}
    \caption{
    Illustration of idle space and time in single-site magic-state cultivation. During the early injection and cultivation stages, only a limited portion of the eventual logical patch is actively used, while the remaining region stays idle until the final escape stage.
    }
    \label{fig:IdleMSC}
\end{figure}

This observation motivates a closer examination of single-site magic-state cultivation. Because the protocol grows one logical state through successive stages within a larger target patch, the early injection and cultivation stages occupy only a limited effective region of the eventual logical footprint~\cite{gidney_magic_2024}. As a result, a nontrivial fraction of the available spacetime volume remains underutilized during the early part of the protocol. Moreover, when a detector event triggers rejection in an early stage, the entire attempt must be restarted, thereby increasing the expected number of attempts required to obtain one acceptable output state. This feature is illustrated in Fig.~\ref{fig:IdleMSC}, where the early stages use only a limited portion of the eventual logical patch while the remaining region stays idle until the final escape stage. Together, these two features define a natural target for further improvement.

In this work, we propose a multiplexed magic-state cultivation scheme that exploits the idle space and time available within a single logical patch during the early stages of cultivation. The key idea is to embed multiple local cultivation opportunities within one patch and to forward a successful outcome to the subsequent escape stage. By increasing the probability of obtaining an acceptable candidate within a single shot, the proposed scheme is designed to reduce the discard rate inherent to postselection-based cultivation and, in turn, to lower the expected number of attempts required to produce one usable logical magic state. In this way, the proposed scheme targets a lower end-to-end resource cost for magic-state cultivation.

\begin{figure*}[!t] 
    \centering
    \centering{\includegraphics[width=17.5cm]{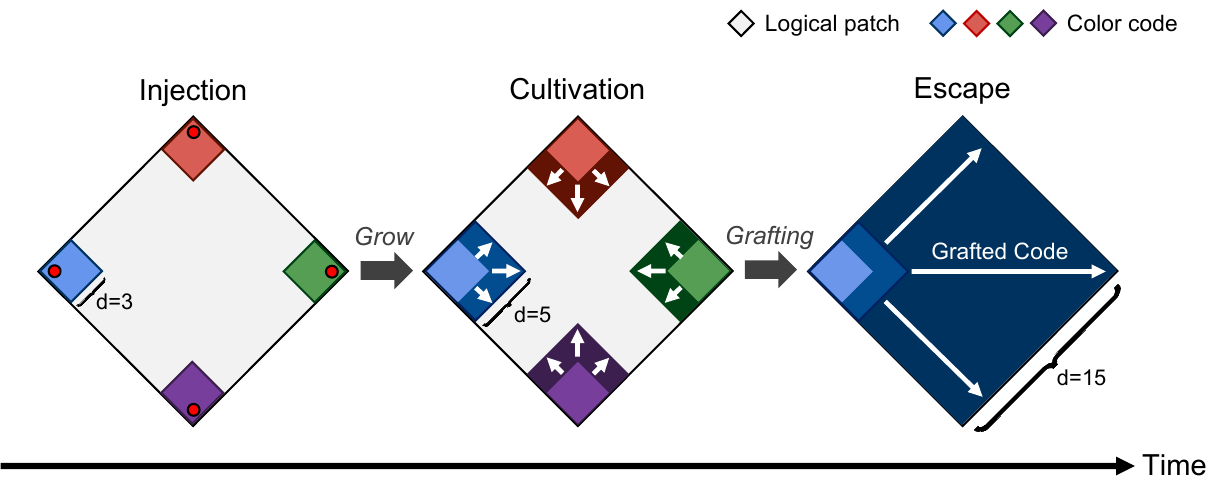}}
    \caption{
    Conceptual illustration of the proposed in-patch multiplexed magic-state cultivation protocol. During injection, four local color-code regions are initialized within a common logical patch. During cultivation, the local regions are grown from \(d=3\) to \(d=5\) and checked in parallel. After a surviving candidate is selected, the protocol returns to the standard single escape path through grafting, producing the final \(d=15\) grafted-code logical patch. Thus, multiplexing is applied only to the early injection-and-cultivation stages, while the final escape stage remains a single continuation path.
    }
    \label{fig:Overview}
\end{figure*}

The main contributions of this paper are summarized as follows:

\begin{itemize}
    \item In-patch multiplexed magic-state cultivation:
    We propose an in-patch multiplexing principle for magic state cultivation(MSC) and instantiate it with a four-site layout that embeds multiple local cultivation opportunities within a single logical patch by using idle space and time during the early injection-and-cultivation stages, while preserving the standard single-path escape stage.

    \item Early-stage postselection-overhead reduction:
    We show that multiplexing changes the early-stage success condition from requiring one predetermined trajectory to pass to requiring at least one of four local trajectories to pass. Under the independent-site approximation, this gives \(D_{\mathrm{IC}}^{(4)} \simeq (D_{\mathrm{IC}}^{(1)})^4\), thereby reducing the early shot-discard rate before escape.

    \item Quantitative stage-level and end-to-end evaluation:
    Through stage-level and full-cycle demonstrations across different local cultivation distances and physical error rates, we evaluate how the proposed protocol reduces the expected-attempt cost relative to the single-site MSC baseline. These results show that in-patch multiplexing can exploit early-stage resource underutilization and reduce expected-attempt overhead while preserving the standard MSC framework.
\end{itemize}

The remainder of this paper is organized as follows.
Section~\ref{sec:Protocol Design} presents the proposed multiplexed magic-state cultivation protocol and its operating principle. Section~\ref{sec:Results} describes the demonstration setup and evaluates the proposed protocol at both the injection-and-cultivation stage and the end-to-end level. Section~\ref{sec:Discussion} discusses the implications, modeling assumptions, and possible extensions of the proposed approach. Finally, Sec.~\ref{sec:Conclusion} concludes the paper. Additional protocol details and supplementary results are provided in the Appendices.

\section{~\label{sec:Protocol Design}Protocol Design}
This section presents the proposed multiplexed magic-state cultivation protocol. We first describe the design motivation, then explain the multiplexed injection-and-cultivation scheme within a single logical patch, and finally describe the end-to-end protocol leading to the escape stage.

\subsection{Overview of the proposed multiplexed MSC}

The proposed protocol extends conventional single-site magic-state cultivation by allowing multiple local cultivation opportunities to coexist within a single logical patch during the early stages of the process. In the standard single-site protocol, a single logical \(|T\rangle\) state is grown through a staged sequence of injection, cultivation, and escape. A more detailed review of the conventional MSC pipeline is provided in Appendix~\ref{sec:Magic State Cultivation}. The proposed multiplexed structure preserves this overall growth paradigm, but modifies the early part of the protocol so that multiple local candidate trajectories can be initiated in parallel before the final escape stage is reached.

Fig.~\ref{fig:Overview} illustrates the conceptual structure of the proposed protocol. In the first stage, injection is performed in parallel at multiple local sites, so that several local \(|T\rangle\) states are initialized within a single logical patch. This makes use of the idle space that would otherwise remain unused in conventional single-site injection. In the second stage, each initialized local state undergoes cultivation within its local region. If an error-detection event occurs at a given site during cultivation, that site is no longer advanced and is instead reset or returned to an inactive state, while the remaining sites continue to evolve. Through this process, the local sites that survive cultivation without triggering an error-detection event form the candidate set for continuation. 

Finally, one successful candidate is forwarded to the escape stage, after which the protocol resumes the standard growth path toward a high-fidelity logical magic state. In the underlying MSC framework, this transition involves grafting from the local color-code growth region to the larger escaped logical patch. The structural distinction between the code patches and the resulting need for a grafted-code transition is discussed in Appendix~\ref{sec:Patch Structures and the Need for Grafting}. In this way, the proposed protocol preserves the staged structure of cultivation while multiplexing the early injection-and-cultivation stages within a single shot.

In the proposed protocol, multiplexing is applied only to the early injection-and-cultivation stages. In this regime, the idle space within a single logical patch is used to host multiple local cultivation sites, whereas the escape stage remains a single continuation path acting on one selected candidate. The purpose of this design is to increase the probability that at least one acceptable local outcome is available before escape and, in turn, to lower the expected number of attempts required to obtain one usable logical magic state.

\subsection{Multiplexed injection and cultivation within a single logical patch}

\begin{figure}[!t]
    \centering{\includegraphics[width=0.455\textwidth]{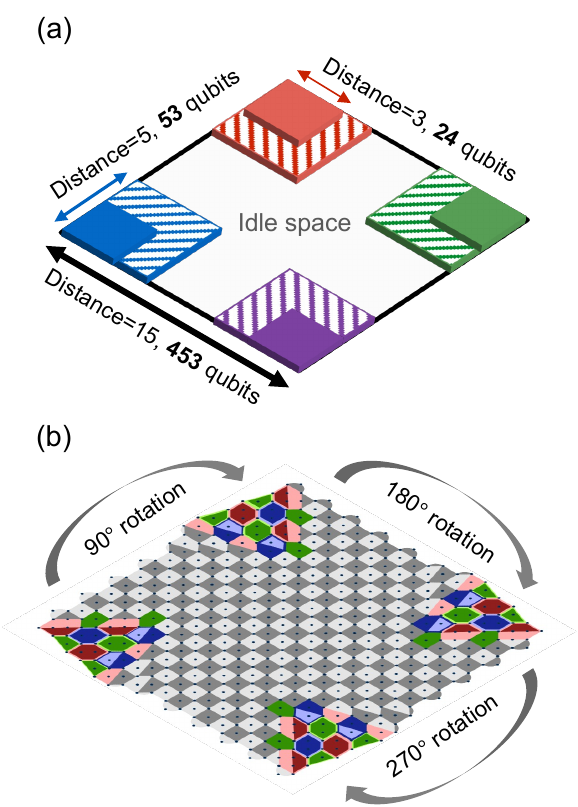}}
    \caption{
    Multiplexed injection-and-cultivation layout within a single logical patch.
    (a) Conceptual placement of the four local cultivation sites in corner-localized regions of a distance-15 logical patch. Each site begins as a distance-3 color-code injection region (24 physical qubits) and is subsequently grown to a distance-5 local cultivation region (53 physical qubits), while the remaining central area forms an idle region during the early stages of the protocol.
    (b) Explicit two-dimensional embedding of the four local code patches within the distance-15 logical patch. The four sites implement the same local growth primitive and are arranged as symmetry-related placements with \(90^\circ\), \(180^\circ\), and \(270^\circ\) in-plane rotations.
    }
    \label{fig:PossibilityandAllocation}
\end{figure}

We describe how the multiplexed early-stage growth is implemented within a single logical patch. As shown in Fig.~\ref{fig:PossibilityandAllocation}(a), the proposed structure places four local cultivation sites in corner-localized regions of a common logical patch. Each local site begins as a distance-3 color-code injection region and is subsequently grown into a distance-5 local cultivation region, while the full logical patch is configured to support the final distance-15 escaped output. In the realization considered here, the distance-3 and distance-5 color-code regions require 24 and 53 physical qubits, respectively. Therefore, given that the final logical patch is sized for the distance-15 escaped state, a substantial idle region naturally remains within the patch during the early stages of the protocol even after the four local growth regions are embedded. This shows that multiple local cultivation opportunities can be accommodated simultaneously within a single logical patch.

\begin{figure*}[!t]
    \centering
    \centering{\includegraphics[width=17.5cm]{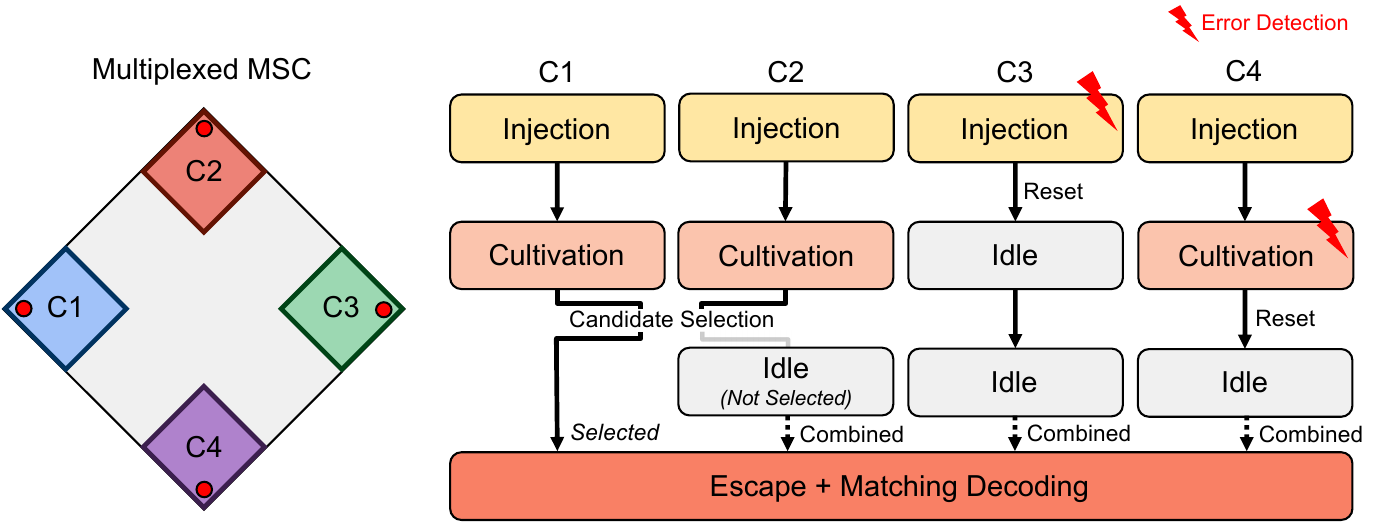}}
    \caption{
    End-to-end flow of the proposed multiplexed MSC protocol. Four local sites (\(C_1\)–\(C_4\)) begin injection and cultivation in parallel within a single logical patch. Sites that trigger an error-detection event are reset and removed from further advancement, while the surviving sites form the candidate set for continuation. One candidate is selected for the final escape stage, and the remaining sites stay idle for the rest of the shot.
    }
    \label{fig:EndtoEnd}
\end{figure*}

The embeddings of these local regions are shown in Fig.~\ref{fig:PossibilityandAllocation}(b). Although the four sites implement the same local preparation and growth primitive, they are placed with different orientations determined by the surrounding patch-boundary geometry and by the requirement that a selected local trajectory be connected consistently to the subsequent grafted-code escape stage. More specifically, the four embeddings can be understood as symmetry-related placements obtained by rotating a common reference local layout in the plane by \(0^\circ\), \(90^\circ\), \(180^\circ\), and \(270^\circ\). This placement preserves the same local operational role at each site while ensuring that the four local regions remain nonoverlapping within the common patch. Moreover, this rotation-related placement allows the standard MSC local growth rule to be implemented in the same manner at each corner allocation. A more explicit two-dimensional lattice embedding within the distance-15 logical patch, together with the detailed geometric configuration of the local regions and the realization of the local growth rule under this rotation pattern, is provided in Appendix~\ref{sec:Possibility and allocation of multiplexed Injection and cultivation}.

\subsection{\label{sec:End-to-end protocol}End-to-end protocol}

Fig.~\ref{fig:EndtoEnd} illustrates the end-to-end execution flow of the proposed multiplexed MSC protocol. The four local sites, denoted by \(C_1,\dots, C_4\), are embedded within a common logical patch and begin the protocol in parallel within a single shot. Injection is first performed site-wise at each local site. If an error-detection event is triggered at a given site during this stage, that site is no longer advanced and is instead returned to an inactive state. Only the sites that complete injection without triggering an error-detection event proceed to cultivation.

The surviving local sites then undergo cultivation, again in parallel. As in the injection stage, each site evolves as a separate local trial. That is, if an error-detection event is detected at one site during cultivation, only that site is removed from further advancement, while the remaining sites continue through cultivation. Consequently, a single shot contains multiple simultaneous local cultivation attempts, but the success or failure of each site is evaluated site-wise. The local sites that complete cultivation without triggering an error-detection event form the candidate set for continuation.

The next step is candidate selection. If no surviving local site completes cultivation, the shot is discarded, and the protocol must be restarted from a new shot. If one or more sites survive, only one of them is selected and forwarded to the subsequent escape stage, while the remaining surviving local sites are not advanced further and remain idle for the remainder of the shot. This selection is justified by both the single-path structure of the standard MSC escape stage and the symmetry of the surviving candidates. Standard MSC prepares one cultivated logical state and then forwards that state through a single escape path, while parallel attempts in the early stages may be viewed as an extension for reducing retry overhead. Moreover, the surviving local sites in the proposed protocol do not correspond to distinct output types, but to symmetry-equivalent candidates preparing the same logical target. Each site is based on a common local preparation-and-growth primitive, while being embedded with a different orientation so as to satisfy the boundary geometry of its assigned corner, the associated local stabilizer support, and topological compatibility with the subsequent grafted-code escape stage. Therefore, even when more than one site survives cultivation, selecting one of them leaves the logical target unchanged and uniquely determines the continuation path leading to the final escaped output.

Once a candidate has been selected, the protocol returns to the standard MSC escape framework. Thus, multiplexing is applied only to the early injection-and-cultivation stages, whereas the final escape stage remains a single continuation path acting on one selected local trajectory. More detailed implementation rules for detector-based candidate formation, site selection, and the treatment of failed and unselected sites are provided in Appendix~\ref{app:selective_escape_detail}.

\section{~\label{sec:Results}Results}

In this section, we evaluate the proposed multiplexed MSC protocol under the same end-to-end demonstration setting used for the standard MSC baseline. We first describe the demonstration setup, including the common stage configurations and evaluation conditions used for comparison. We then examine how multiplexing changes the discard behavior during the injection-and-cultivation stages and, finally, analyze how this change translates into the full-cycle cost of obtaining one escaped logical output.

\subsection{~\label{sec:Demonstration setup}Demonstration setup}

We evaluate the proposed multiplexed MSC protocol under the same demonstration setting used for the standard single-site MSC baseline. The purpose of this evaluation is to isolate the effect of multiplexing the early injection-and-cultivation stages while keeping the final escape stage and the decoder-based acceptance procedure unchanged. Our demonstration is based on the public code and data artifact accompanying the standard MSC study~\cite{gidney_magic_2024, gidney_data_2024}. Starting from the baseline implementation, we extend the early injection-and-cultivation stage to support four site-wise local cultivation trials within a single logical patch, while preserving the original escape-stage circuit construction, decoding pipeline, and gap-based acceptance procedure for the selected candidate.

The main demonstration parameters are summarized in Table~\ref{tab:demo_setup}. All demonstrations use the $Y$-basis setting, CSS codes, unitary injection, and a uniform depolarizing noise model~\cite{gidney_magic_2024, PhysRevA.83.020302, PhysRevA.90.062320}. For the full-cycle evaluation, the escape-stage syndrome is decoded using 
the same desaturation decoder for both the single-site baseline and the multiplexed protocol~\cite{gidney_magic_2024,hao_compilation_2025, chase_clifft_2026}.

We compare the single-site MSC baseline with the multiplexed protocol defined in Sec.~\ref{sec:End-to-end protocol}. In the baseline, an early detector event discards the entire shot. In the multiplexed protocol, early detector events are handled site-wise, and the shot is discarded only if all four local sites fail before escape. Candidate selection is performed before escape and does not use post-escape decoder-gap information.

We evaluate the protocol at two levels. At the injection-and-cultivation level, we measure the early-stage discard rate \(D_{\mathrm{IC}}\), defined as the fraction of shots discarded before escape, and the corresponding expected attempts \(A_{\mathrm{IC}}\) required to obtain one accepted early-stage candidate. At the full-cycle level, we include injection, cultivation, escape, and decoding, and measure the expected attempts per kept logical output \(A_{\mathrm{full}}\) and the logical error rate per kept shot \(p_L\). The full-cycle analysis compares the trade-off between \(A_{\mathrm{full}}\) and \(p_L\) as the escape-stage gap threshold is varied. A more detailed definition and interpretation of the gap-based acceptance metrics are provided in Appendix~\ref{sec:Gap-based acceptance and distance-dependent logical-error behavior}.

\begin{table}
\caption{Demonstration parameters}
\label{tab:demo_setup}
\centering
\renewcommand{\arraystretch}{1.12}
\setlength{\tabcolsep}{5pt}
\begin{tabular}{ll}
\hline
\textbf{Symbol} & \textbf{Definition} \\
\hline
$d_1$ &
\parbox[t]{6.0cm}{\justifying\setlength{\parindent}{0pt}\noindent Local cultivation distance used for the injection-and-cultivation stages; $d_1 \in \{3,5\}$.} \\

$d_2$ &
\parbox[t]{6.0cm}{\justifying\setlength{\parindent}{0pt}\noindent Final escaped patch distance used for the logical output; $d_2=15$.} \\

$r_1$ &
\parbox[t]{6.0cm}{\justifying\setlength{\parindent}{0pt}\noindent Number of growing rounds in the local cultivation stage; $r_1=d_1$.} \\

$r_2$ &
\parbox[t]{6.0cm}{\justifying\setlength{\parindent}{0pt}\noindent Number of rounds used in the final escape stage; $r_2=5$.} \\

$p$ &
\parbox[t]{6.0cm}{\justifying\setlength{\parindent}{0pt}\noindent Physical error rate of the uniform depolarizing noise model; $p \in \{5\times10^{-4},\,10^{-3},\,2\times10^{-3}\}$.} \\
\hline
\end{tabular}
\end{table}

\subsection{~\label{sec:Injection and Cultivation Discard rate}Injection-and-cultivation discard rate}

We begin by isolating the injection and cultivation stages from the full MSC cycle in order to evaluate the direct effect of multiplexing on early-stage discard behavior. This analysis focuses on the part of the protocol where multiplexing is actually applied, before the selected candidate is forwarded to the escape stage. In the single-site baseline,
a detector event during injection or cultivation rejects the only local trajectory, and therefore, the entire shot is discarded. In contrast, in the multiplexed protocol, the same detector-based rejection rule is applied site-wise. A detector event removes only the corresponding site from further advancement, while the remaining sites continue through the early stages. Therefore, in the multiplexed protocol, the entire shot is discarded only when all four local sites fail during injection and cultivation.

\begin{figure*}[!t]
    \centering
    \centering{\includegraphics[width=17.5cm]{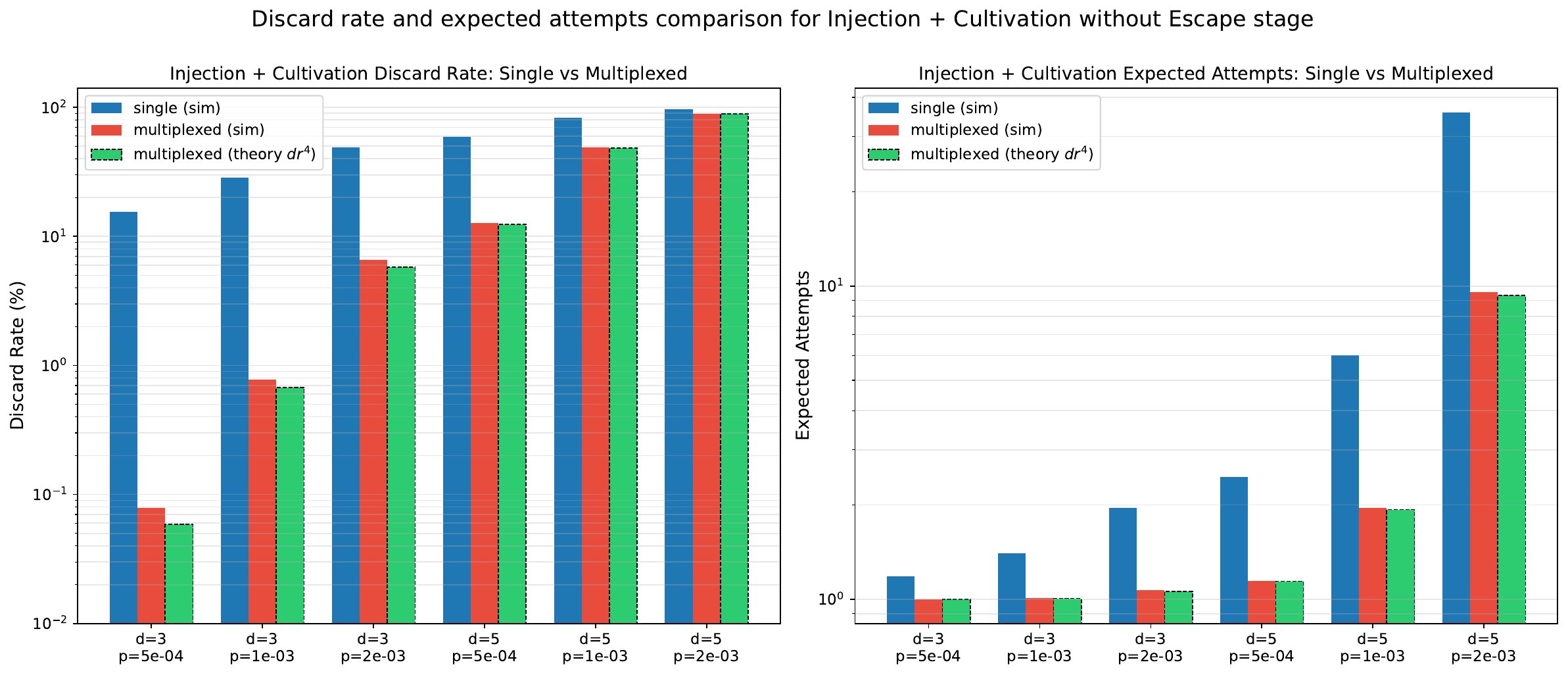}}
    \caption{
    Injection-and-cultivation-stage discard behavior before escape. 
    The left panel compares the discard rate, and the right panel compares the corresponding expected attempts for the single-site baseline and the four-site multiplexed protocol. 
    Blue and red bars denote direct demonstration results for the single-site and multiplexed cases, respectively, while green hatched bars show the independent-site estimate \(D_{\mathrm{IC}}^{(4)}=(D_{\mathrm{IC}}^{(1)})^4\).
    }
    \label{fig:DiscardAndAttempts}
\end{figure*}

\begin{table*}[t]
\caption{
Injection-and-cultivation discard rate and expected-attempt reduction before the escape stage.
}
\label{tab:ic_attempt_reduction}
\begin{ruledtabular}
\begin{tabular}{ccccccc}
\(d_1\) & \(p\) 
& \(D_{\mathrm{IC}}^{(1)}\) 
& \(D_{\mathrm{IC}}^{(4)}\) 
& \(A_{\mathrm{IC}}^{(1)}\) 
& \(A_{\mathrm{IC}}^{(4)}\) 
& \(\rho_{\mathrm{IC}}\) \\
\hline
3 & \(5\times10^{-4}\) & \(15.60\%\) & \(0.06\%\) & 1.1849 & 1.0006 & \(15.55\%\) \\
3 & \(1\times10^{-3}\) & \(28.73\%\) & \(0.76\%\) & 1.4031 & 1.0077 & \(28.18\%\) \\
3 & \(2\times10^{-3}\) & \(49.03\%\) & \(6.56\%\) & 1.9620 & 1.0702 & \(45.46\%\) \\
5 & \(5\times10^{-4}\) & \(59.31\%\) & \(12.69\%\) & 2.4575 & 1.1453 & \(53.40\%\) \\
5 & \(1\times10^{-3}\) & \(83.44\%\) & \(48.97\%\) & 6.0397 & 1.9595 & \(67.56\%\) \\
5 & \(2\times10^{-3}\) & \(97.20\%\) & \(89.68\%\) & 35.7590 & 9.6876 & \(72.91\%\) \\
\end{tabular}
\end{ruledtabular}
\end{table*}

Let \(D_{\mathrm{IC}}^{(1)}\) denote the discard probability of a single local injection-and-cultivation trajectory. The corresponding early-stage pass probability is
\[
P_{\mathrm{IC}}^{(1)} = 1-D_{\mathrm{IC}}^{(1)} .
\]
In the four-site multiplexed protocol, an early-stage candidate is available if at least one of the four local sites passes both injection and cultivation. Under the ideal independent-site approximation, the four-site early-stage pass probability is therefore
\[
P_{\mathrm{IC}}^{(4)}
=
1-\left(D_{\mathrm{IC}}^{(1)}\right)^4 ,
\]
and the corresponding discard probability is
\[
D_{\mathrm{IC}}^{(4)}
=
\left(D_{\mathrm{IC}}^{(1)}\right)^4 .
\]
Thus, multiplexing changes the early-stage acceptance condition from requiring one specific trajectory to pass to requiring at least one of the four local trajectories to pass. The expected number of attempts required to obtain one accepted early-stage candidate is
\[
A_{\mathrm{IC}}^{(n)}
=
\frac{1}{1-D_{\mathrm{IC}}^{(n)}} ,
\]
where \(n=1\) denotes the single-site baseline and \(n=4\) denotes the four-site multiplexed protocol. We quantify the relative reduction in expected attempts before escape as
\[
\rho_{\mathrm{IC}}
=
\left(
1-\frac{A_{\mathrm{IC}}^{(4)}}{A_{\mathrm{IC}}^{(1)}}
\right)\times 100\% .
\]

Figure~\ref{fig:DiscardAndAttempts} compares the injection-and-cultivation discard rate and the corresponding expected attempts. For \(d_1=3\), the
single-site discard rate ranges from \(15.60\%\) to \(49.03\%\) over the evaluated physical error rates, whereas the four-site multiplexed protocol
reduces it to \(0.06\%\)--\(6.56\%\). This reduction keeps the corresponding expected attempts close to one, decreasing \(A_{\mathrm{IC}}\) from
\(1.1849\)--\(1.9620\) in the baseline to \(1.0006\)--\(1.0702\) in the multiplexed protocol.

For \(d_1=5\), the larger local cultivation region leads to higher single-site discard rates, ranging from \(59.31\%\) to \(97.20\%\). The multiplexed protocol reduces these values to \(12.69\%\)--\(89.68\%\). As a result, the expected attempts decrease from \(2.4575\)--\(35.7590\) in the single-site baseline to \(1.1453\)--\(9.6876\) in the multiplexed protocol. The largest reduction occurs at \(p=2\times10^{-3}\), where the expected attempts are reduced by \(45.46\%\) for \(d_1=3\) and by \(72.91\%\) for \(d_1=5\). The full set of values is summarized in Table~\ref{tab:ic_attempt_reduction}.

The results of the proposed approach closely follow the independent-site estimate \(D_{\mathrm{IC}}^{(4)}=(D_{\mathrm{IC}}^{(1)})^4\), indicating that the reduction in the early-stage discard rate is well explained by the probability of obtaining at least one successful local candidate.

\begin{figure*}[!t] 
    \centering
    \centering{\includegraphics[width=17.5cm]{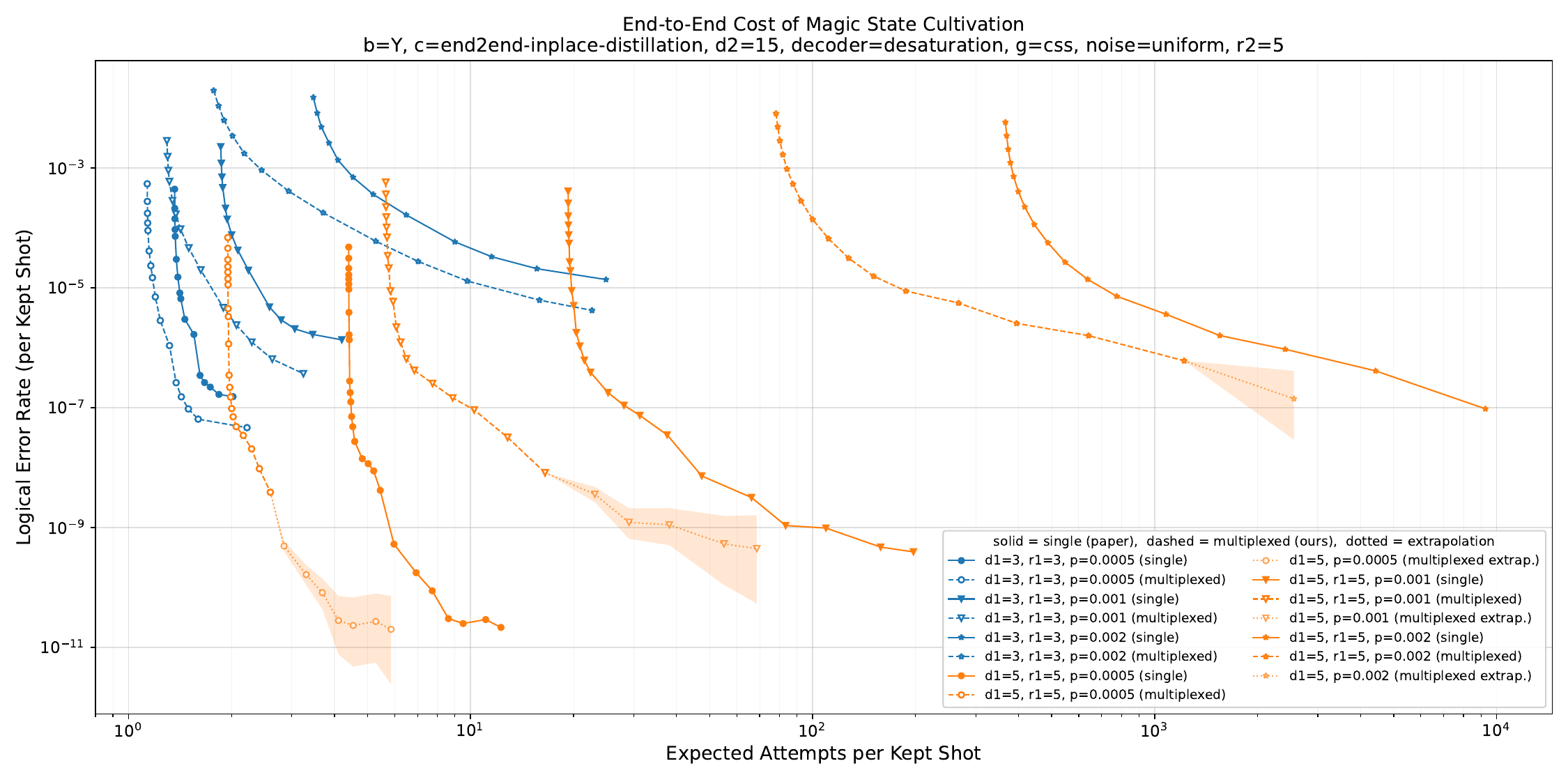}}
    \caption{
    End-to-end cost comparison of the standard single-site MSC and the proposed multiplexed MSC protocol. The horizontal axis shows the expected attempts per kept shot, and the vertical axis shows the logical error rate per kept shot after escape-stage decoding and gap-based acceptance. Each curve is obtained by sweeping the gap threshold \(G\); solid curves denote the single-site baseline, and dashed curves denote the multiplexed protocol. Dotted segments and shaded regions indicate extrapolated estimates in the large-threshold regime, where the number of remaining error samples is limited. Across the evaluated settings, the multiplexed protocol mainly shifts the curves toward smaller expected attempts.
    }
    \label{fig:E2ECost}
\end{figure*}

\subsection{~\label{sec:End-to-end Cost}End-to-end cost}

We evaluate how the proposed multiplexed MSC protocol changes the overall cost required to obtain one kept logical magic state. As shown in the previous subsection, the direct effect of multiplexing is to reduce discards during the early injection and cultivation stages. However, the final output of MSC is not determined by early-stage survival alone. A candidate that passes injection and cultivation must be forwarded to the escape stage and must also satisfy the escape-stage decoding and gap-based acceptance criterion before it is counted as a kept shot. We therefore compare the expected attempts per kept shot and the logical error rate from a full-cycle perspective, including injection, cultivation, escape, and decoding.

Figure~\ref{fig:E2ECost} compares the end-to-end cost of the standard single-site MSC baseline and the proposed multiplexed protocol. The horizontal axis shows the expected number of attempts required to obtain one kept shot, while the vertical axis shows the logical error rate per kept shot. Solid curves correspond to the single-site MSC baseline, and dashed curves correspond to the multiplexed protocol. Each curve is obtained by varying the gap threshold of the escape-stage decoder. As the gap threshold is increased, shots for which the decoder has a smaller separation between the two logical hypotheses are additionally discarded. Consequently, the expected number of attempts generally increases, while the logical error rate tends to decrease.

To quantify the full-cycle reduction in expected attempts, we define
\[
\rho_{\mathrm{full}}
=
\left(
1-\frac{A_{\mathrm{full}}^{(4)}}{A_{\mathrm{full}}^{(1)}}
\right)\times 100\% ,
\]
where \(A_{\mathrm{full}}^{(1)}\) and \(A_{\mathrm{full}}^{(4)}\) denote the expected attempts per kept logical output for the single-site baseline and the four-site multiplexed protocol, respectively.

\begin{table}[t]
\caption{
Full-cycle expected-attempt reduction at \(G=0\), including injection, cultivation, escape, and decoding. The values correspond to the \(G=0\) operating point of the gap-threshold sweep in Fig.~\ref{fig:E2ECost}.}
\label{tab:full_attempt_reduction}
\begin{ruledtabular}
\begin{tabular}{ccccc}
\(d_1\) & \(p\)
& \(A_{\mathrm{full}}^{(1)}\)
& \(A_{\mathrm{full}}^{(4)}\)
& \(\rho_{\mathrm{full}}\) \\
\hline
3 & \(5\times10^{-4}\) & 1.3632 & 1.1332 & \(16.87\%\) \\
3 & \(1\times10^{-3}\) & 1.8562 & 1.2911 & \(30.44\%\) \\
3 & \(2\times10^{-3}\) & 3.4288 & 1.7473 & \(49.04\%\) \\
5 & \(5\times10^{-4}\) & 4.4014 & 1.9503 & \(55.69\%\) \\
5 & \(1\times10^{-3}\) & 19.2822 & 5.6465 & \(70.72\%\) \\
5 & \(2\times10^{-3}\) & 364.7853 & 77.7446 & \(78.69\%\) \\
\end{tabular}
\end{ruledtabular}
\end{table}

Across the evaluated parameter settings, the multiplexed protocol reduces the expected number of attempts required to obtain one kept logical output. As summarized in Table~\ref{tab:full_attempt_reduction}, the full-cycle expected attempts are reduced by \(16.87\%\)--\(78.69\%\) relative to the single-site MSC baseline. The reduction becomes more pronounced as the physical error rate and local cultivation distance increase. At \(p=2\times10^{-3}\), the expected attempts decrease from \(3.4288\) to \(1.7473\) for \(d_1=3\), corresponding to a \(49.04\%\) reduction, and from \(364.7853\) to \(77.7446\) for \(d_1=5\), corresponding to a \(78.69\%\) reduction.

This full-cycle reduction originates from the same early-stage mechanism identified in the injection-and-cultivation analysis. In the single-site protocol, a detector event during injection or cultivation rejects the only local trajectory and therefore discards the whole shot. In contrast, the proposed protocol runs four local cultivation sites in parallel within one shot, and the protocol can proceed to escape if at least one of these sites completes injection and cultivation without triggering a detector event. If \(p_{\mathrm{pass}}\) denotes the probability that a single local trajectory passes the early stages, then, under the ideal independent-site approximation, the probability that at least one of the four sites passes is
\begin{equation}
P_{\mathrm{pass}}^{(4)}
=
1-\left(1-p_{\mathrm{pass}}\right)^4 .
\end{equation}
This increased probability of obtaining an early-stage candidate reduces the number of attempts required to obtain a kept output, which is why the dashed curves in Fig.~\ref{fig:E2ECost} are shifted toward smaller expected attempts compared with the solid curves.

The logical error rate should be interpreted together with the escape-stage gap threshold. The primary role of multiplexing is not to uniformly reduce the logical error rate of each accepted output, but to reduce early-stage discards and thereby lower the expected attempts required to obtain kept outputs. At small gap thresholds, shots with only a small separation between the two decoder hypotheses can still be included among the kept shots. As a result, the logical error rate of the multiplexed protocol can be comparable to, or in some cases larger than, that of the single-site baseline. This occurs because the absence of detector events during injection and cultivation does not by itself guarantee that the selected candidate will have a large decoder gap after escape.

As the gap threshold is increased, shots with smaller gap values are removed first, and the final kept set is restricted to cases in which the decoder makes a more separated decision. In Fig.~\ref{fig:E2ECost}, this gap-dependent behavior depends on the local cultivation distance. For the \(d_1=3\) cases, the multiplexed protocol exhibits regions at larger gap thresholds where the logical error rate is lower than that of the single-site MSC baseline. For the \(d_1=5\) cases, by contrast, the benefit of multiplexing appears more clearly as a reduction in the expected number of attempts rather than as a uniform reduction in logical error rate. Thus, the proposed multiplexing lowers the expected-attempt cost required to obtain a kept output, while the logical-error behavior remains governed by the local cultivation distance and the escape-stage gap threshold. A detailed analysis of the threshold-dependent correct/error composition and its distance dependence is provided in Appendix~\ref{sec:Gap-based acceptance and distance-dependent logical-error behavior}.

\section{~\label{sec:Discussion}Discussion}

The results show that multiplexing can reduce the postselection loss in the early injection-and-cultivation stages of magic-state cultivation. The proposed protocol places multiple local cultivation opportunities within a single logical patch and forwards one candidate that passes the early stages to the standard escape pathway. The final escape stage and the decoder-based acceptance procedure are kept identical to those of the standard MSC framework. Therefore, the main benefit of the proposed method comes from increasing the probability of obtaining an acceptable candidate before escape, which reduces the cost associated with repeated restarts.

In the demonstration in our work, the four local sites are placed in nonoverlapping regions, and each site uses the same local cultivation primitive. We also apply the same uniform depolarizing noise model to gate, measurement, and idle operations, with operation-level errors sampled independently, following common stochastic-noise assumptions used in quantum-error-correction works~\cite{PhysRevA.90.062320, PhysRevX.13.031007, PhysRevX.2.021004, 10.5555/2011362.2011368}. Under these modeling conditions, the site-wise injection-and-cultivation failures can be approximated as independent and identically distributed (i.i.d.) Bernoulli events~\cite{jones_low-overhead_2013, tiurev_correcting_2023, cruz_quantum_2023}. If \(D_{\mathrm{IC}}^{(1)}\) denotes the discard probability of a single local trajectory, the four-site protocol discards the entire shot only when all four local sites fail during the early stages. Therefore,
\[
D_{\mathrm{IC}}^{(4)}
=
\left(D_{\mathrm{IC}}^{(1)}\right)^4 .
\]
The agreement between the demonstration results in Fig.~\ref{fig:DiscardAndAttempts} and this independent-site prediction indicates that, under the four-corner layout and the uniform depolarizing noise model with idle noise, the early-stage discard reduction is consistently described by the i.i.d. approximation. In this setting, where the local sites are nonoverlapping and subject to the same local circuit primitive and noise conditions, the multiplexing effect is quantified by the probability that at least one of the four local trials succeeds.

In real hardware, effective independence between local sites may not be fully maintained even when the sites are spatially separated~\cite{harper_efficient_2020,sarovar_detecting_2020,rudinger_experimental_2021}. 
In an actual device, common control lines, shared readout resources, simultaneous operations, or hardware-level crosstalk may introduce correlations between failure events at different local sites~\cite{gambetta_characterization_2012,PhysRevApplied.14.024042,PhysRevApplied.16.024063,PRXQuantum.2.040313}. 
The focus of this work, however, is the architectural efficiency enabled by in-patch multiplexing rather than the detailed crosstalk profile of a specific hardware platform. 
Hardware-specific crosstalk can affect the quantitative magnitude of the realized gain, but it does not alter the fundamental retry-reduction mechanism of the protocol: multiplexing allows the shot to proceed whenever at least one local cultivation site survives before escape. 
Thus, crosstalk should be viewed as a hardware-dependent factor affecting the observed gain, rather than as a change to the underlying architectural benefit of in-patch multiplexing.

This distinction can be expressed more generally in terms of the early-stage discard probability. Let \(\chi_i \in \{0,1\}\) denote the early-stage survival indicator of site \(i\), where \(\chi_i=1\) means that site \(i\) passes injection and cultivation, and \(\chi_i=0\) means that the site fails during the early stages. The discard probability of the multiplexed protocol is then
\[
D_{\mathrm{IC}}^{(4)}
=
\Pr(\chi_1=0,\chi_2=0,\chi_3=0,\chi_4=0).
\]
This expression reduces to \(\left(D_{\mathrm{IC}}^{(1)}\right)^4\) only under the i.i.d. assumption. If the site failure probabilities are different but the failure events remain independent, then, for \(D_i=\Pr(\chi_i=0)\),
\[
D_{\mathrm{IC}}^{(4)}
=
\prod_{i=1}^{4}D_i .
\]
By contrast, positive correlations induced by crosstalk can increase the simultaneous failure probability and lead to a larger discard probability than the independent-site estimate. From this perspective, the i.i.d. model serves as an ideal baseline for interpreting the multiplexing gain.

These considerations suggest several directions for future extensions. 
First, multiplexed MSC can be evaluated using hardware-calibrated noise models that include site-dependent error rates and crosstalk-induced correlations. 
Second, the candidate-selection rule can be extended from predetermined or site-agnostic selection to a rule that incorporates site-dependent reliability. 
In the present protocol, candidate selection is performed before escape and does not use post-escape decoder-gap information. 
In a hardware-aware implementation, calibration data or early-stage detector history could be used to select a more reliable site among the surviving candidates.

A broader extension concerns scaling the in-patch multiplexing principle beyond the specific four-site layout studied in this work. 
The four-site construction should be understood as one concrete use case rather than as a fixed limit of the protocol. 
In fault-tolerant quantum error correction, lowering the logical error rate requires increasing the code distance when the physical error rate is below threshold; recent surface-code experiments have demonstrated this distance-dependent logical-error suppression, and large-scale surface-code architectures organize computation in patch-based spacetime layouts with explicit space-time trade-offs~\cite{acharya_suppressing_2023,acharya_quantum_2025,Litinski2019gameofsurfacecodes}. 
Therefore, as FTQC systems target lower logical error rates and larger workloads, larger escaped patches and more spacious layouts become a natural architectural setting for magic-state preparation. 
In such settings, additional idle regions may be available during the early injection-and-cultivation stages, allowing more local cultivation sites to be embedded and operated in parallel.

This motivates a general \(k\)-site multiplexing extension, where \(k\) is determined by the available patch area, nonoverlap constraints, boundary compatibility with the escape path, and hardware-control constraints. 
If \(P_i\) denotes the early-stage pass probability of site \(i\), then the early-stage pass probability in the independent but nonidentical setting is
\[
P_{\mathrm{pass}}^{(k)}
=
1-\prod_{i=1}^{k}(1-P_i).
\]
In the identical-site case, where \(P_i=P\), this becomes
\[
P_{\mathrm{pass}}^{(k)}
=
1-(1-P)^k,
\qquad
D_{\mathrm{IC}}^{(k)}
=
(1-P)^k
=
\left(D_{\mathrm{IC}}^{(1)}\right)^k .
\]
Under the independent-site approximation, the early-stage discard probability decreases exponentially with the number of multiplexed sites \(k\), while the probability of obtaining at least one candidate increases accordingly. 
The realized benefit, however, must be balanced against layout feasibility, control complexity, crosstalk management, and reset and idle treatment.

\section{~\label{sec:Conclusion}Conclusion}

In this work, we proposed an in-patch multiplexing scheme for reducing the postselection overhead of magic-state cultivation. The proposed protocol uses idle early-stage resources within a single logical patch to create multiple local cultivation opportunities in parallel, while keeping the final escape stage and decoder-based acceptance procedure identical to those of the single-site MSC baseline. The injection-and-cultivation analysis showed that multiplexing directly reduces early-stage postselection loss; under the independent-site approximation, the four-site discard probability follows \((D_{\mathrm{IC}}^{(1)})^4\), and the four-site demonstration results closely matched this prediction. Quantitatively, the proposed protocol reduces the injection-and-cultivation expected attempts by up to \(72.91\%\) and the full-cycle expected attempts per kept logical output by up to \(78.69\%\) across the evaluated settings, with the largest reduction observed at \(d_1=5\) and \(p=2\times10^{-3}\). The logical-error behavior after escape remains governed by the gap-based acceptance threshold and the local cultivation distance, indicating that the primary benefit of multiplexing is the reduction of expected attempts rather than a uniform reduction in logical error rate. These results show that in-patch multiplexing can reduce early-stage resource underutilization and expected-attempt overhead while preserving the standard MSC framework.

% In this work, we proposed an in-patch multiplexing scheme that reduces the postselection overhead of magic-state cultivation by using idle early-stage resources within a single logical patch. The proposed protocol creates four local cultivation opportunities in parallel, while keeping the final escape stage and decoder-based acceptance procedure identical to those of the single-site MSC baseline. Under the independent-site approximation, the four-site injection-and-cultivation discard probability follows \((D_{\mathrm{IC}}^{(1)})^4\), which is consistent with the simulation results. Across the evaluated settings, the proposed protocol reduces the injection-and-cultivation expected attempts by up to \(72.91\%\) and the full-cycle expected attempts per kept logical output by up to \(78.69\%\), with the largest reduction observed at \(d_1=5\) and \(p=2\times10^{-3}\). These results show that in-patch multiplexing lowers expected-attempt overhead while preserving the standard MSC framework.

\begin{figure*}[!t]
    \centering{\includegraphics[width=15.5cm]{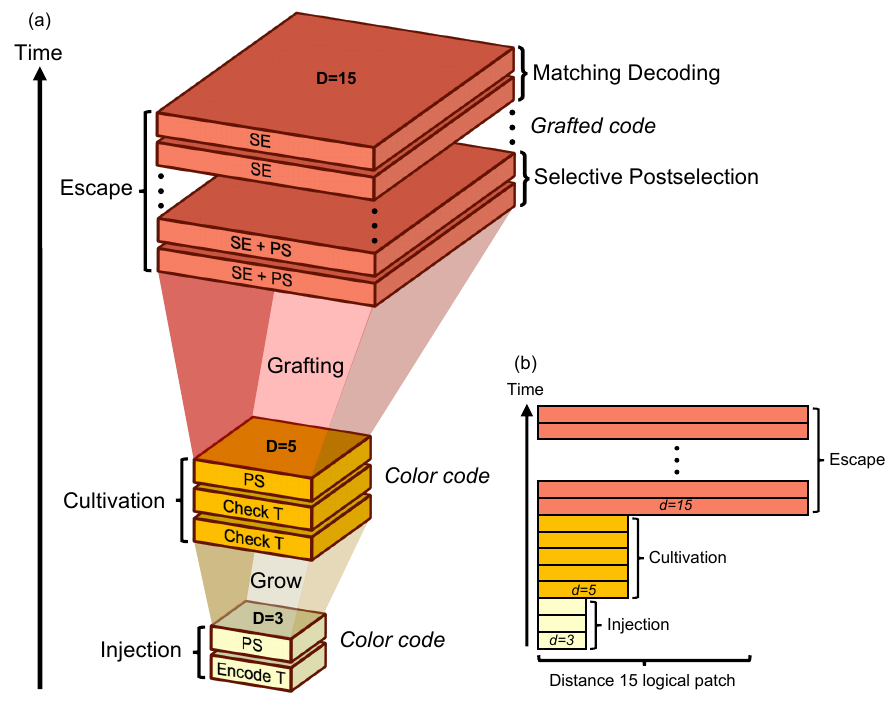}}
    \caption{
    Overview of the standard magic-state cultivation (MSC) pipeline.
    (a) A staged spacetime view of MSC, consisting of an injection stage that initializes a distance-3 color-code logical state, a cultivation stage that grows the state to distance 5 through repeated checking and postselection, and an escape stage that transfers the cultivated state through grafting into a larger distance-15 patch, followed by matching-based decoding.
    (b) The same process viewed in terms of effective patch usage over time, highlighting that the early injection and cultivation stages occupy only a limited portion of the eventual distance-15 logical patch.
    }
    \label{fig:OriginalMSC}
\end{figure*}

% The full-cycle analysis showed that this early-stage discard reduction lowers the expected number of attempts required to obtain one kept logical output. Across the evaluated settings, the multiplexed protocol shifted the cost curves toward smaller expected attempts. The logical-error behavior after escape remained governed by the gap-based acceptance threshold and depended on the local cultivation distance: for \(d_1=3\), the multiplexed protocol showed regimes where both expected attempts and logical error rate were reduced, whereas for \(d_1=5\), the main benefit appeared as reduced expected attempts with comparable gap-dependent logical-error behavior. These results show that in-patch multiplexing can reduce early-stage resource underutilization and lower the retry cost while preserving the standard MSC framework. Future work can extend this framework to hardware-aware settings that account for inter-site crosstalk and correlated failures.

\begin{acknowledgments}
This work was supported in part by the Institute for Information \& communications Technology Planning \& Evaluation (IITP) grant funded by the Korean government (MSIT) (No. RS-2020-II200014, A Technology Development of Quantum OS for Fault-tolerant Logical Qubit Computing Environment); Quantum Science and Technology Flagship Project (Quantum Computing) through the National Research Foundation of Korea(NRF) funded by the Korean government (Ministry of Science and ICT(MSIT)) (No. RS-2025-25464760); and Creation of the quantum information science R\&D ecosystem (based on human resources) through the National Research Foundation of Korea (NRF) funded by the Korean government (Ministry of Science and ICT (MSIT)) (No. RS-2023-00256050).
\end{acknowledgments}

\appendix

\section{\label{sec:preliminaries}Preliminaries}

\subsection{\label{sec:Magic State Cultivation}Magic state cultivation}

Magic-state preparation is a standard route to fault-tolerant non-Clifford computation, with magic-state distillation serving as the dominant framework for producing high-fidelity logical resource states from noisier encoded inputs~\cite{10.5555/2012098.2012105, haah_codes_2018, litinski_magic_2019, surti_efficient_2025}. Within this broader landscape, magic-state cultivation has been proposed as a lower-cost alternative for preparing logical \(|T\rangle\) states in the intermediate logical-error regime. Rather than consuming many encoded magic states in a factory-style purification protocol, MSC grows a single logical \(|T\rangle\) state through a staged sequence of injection, cultivation, and escape~\cite{gpvl-lg4c, hetenyi_constant_2026, wan_simulating_2026}.

Figure~\ref{fig:OriginalMSC} summarizes this standard MSC workflow. During the injection stage, a local logical state is initialized in a small-distance color-code region. In the cultivation stage, the state is grown to a larger intermediate distance through additional checks and postselection. The final escape stage then transfers the cultivated state into a larger logical patch, after which the protocol continues toward a high-fidelity logical output and matching-based decoding. Figure~\ref{fig:OriginalMSC}(b) highlights the same process from the viewpoint of the effective logical-patch footprint over time: the early injection and cultivation stages use only a limited portion of the eventual patch required for the final escaped state. This staged growth pattern is the key structural feature inherited by the present work. The proposed multiplexed protocol does not replace the MSC pipeline with a different preparation mechanism; instead, it modifies how the early stages use the logical patch by embedding multiple local cultivation opportunities in parallel, while preserving the same overall progression toward a single escaped logical output.

\begin{figure}[!t]
    \centering{\includegraphics[width=0.455\textwidth]{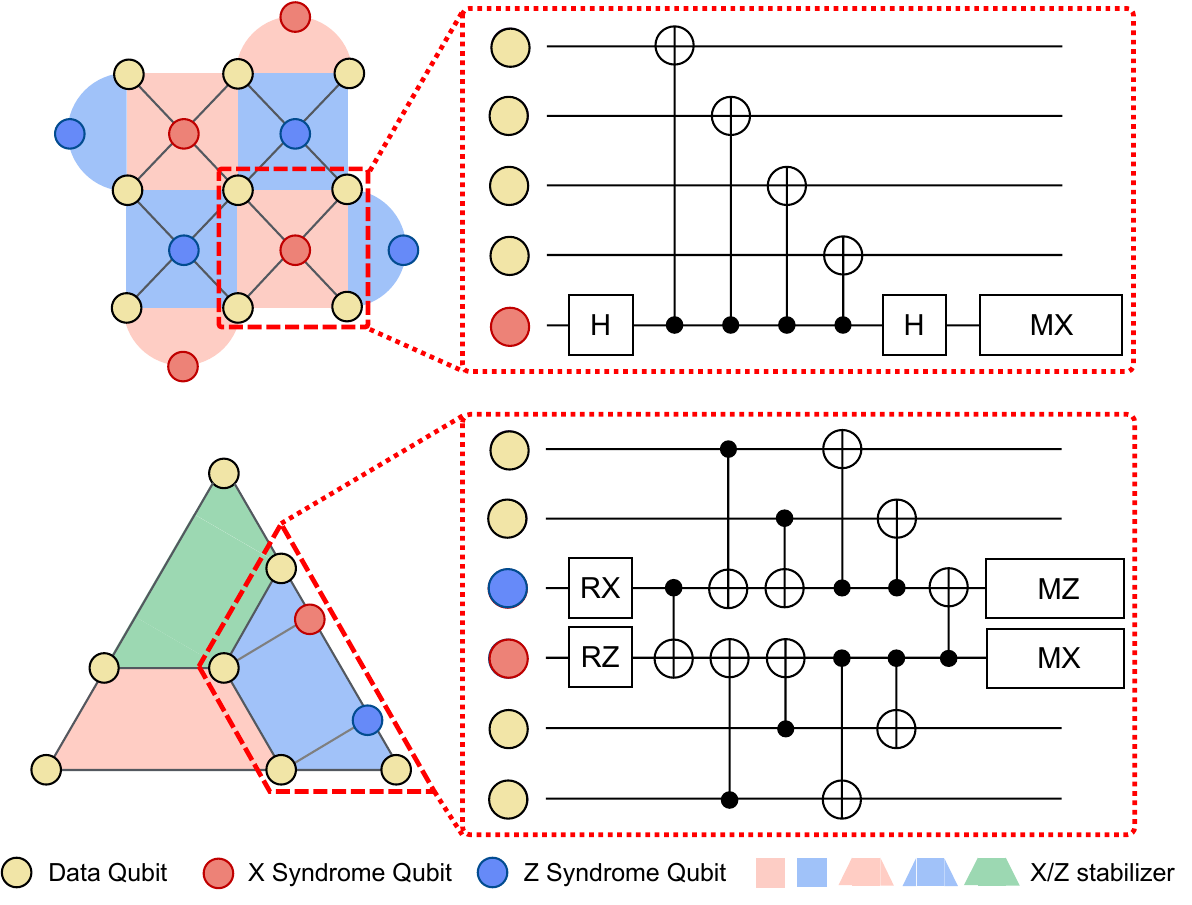}}
    \caption{
    Structural comparison between distance-3 surface-code and color-code patches and their representative stabilizer measurement circuits. The top panel shows a distance-3 surface-code patch with an example stabilizer measurement circuit, while the bottom panel shows a distance-3 color-code patch with the corresponding color-code stabilizer measurement circuit. The difference in stabilizer arrangement and syndrome-extraction structure illustrates why a grafted transition is needed in magic-state cultivation.
    }
    \label{fig:ColorAndSurfaceCode}
\end{figure}

\subsection{\label{sec:Patch Structures and the Need for Grafting}Patch structures and the need for grafting}

The standard MSC protocol combines local growth on color-code regions with final escape into a larger logical patch. These code regions are not structurally identical. Surface-code patches and color-code patches differ in stabilizer geometry and in the corresponding syndrome-extraction circuits~\cite{PhysRevA.86.032324, PhysRevLett.129.030501,krinner_realizing_2022, lacroix_scaling_2025, PRXQuantum.5.030352}. Figure~\ref{fig:ColorAndSurfaceCode} illustrates this distinction schematically. The upper panel shows a representative surface-code stabilizer layout and the associated measurement circuit, while the lower panel shows the corresponding structure for a color-code patch.

Because the local growth region and the final escaped patch are based on different code structures, the cultivated state cannot simply be continued by enlarging the same local patch. Instead, MSC employs a grafted transition that connects the local color-code growth region to the larger escaped logical patch. In this sense, grafting is not an auxiliary detail but a necessary structural step in the MSC pipeline: it mediates the transition from the local color-code region used during injection and cultivation to the larger patch that supports the final escaped output.

This distinction is directly relevant to the main text. In the proposed multiplexed protocol, the parallel local sites are embedded only in the early growth stages, where the local color-code regions remain spatially separated within a single larger patch. Once a surviving site is selected, the protocol returns to the same escape pathway as in standard MSC, including the grafted transition to the final logical region.

\section{\label{sec:Possibility and allocation of multiplexed Injection and cultivation}Geometric feasibility and allocation of multiplexed injection and cultivation}

\begin{figure*}[!t]
    \centering{\includegraphics[width=10.5cm]{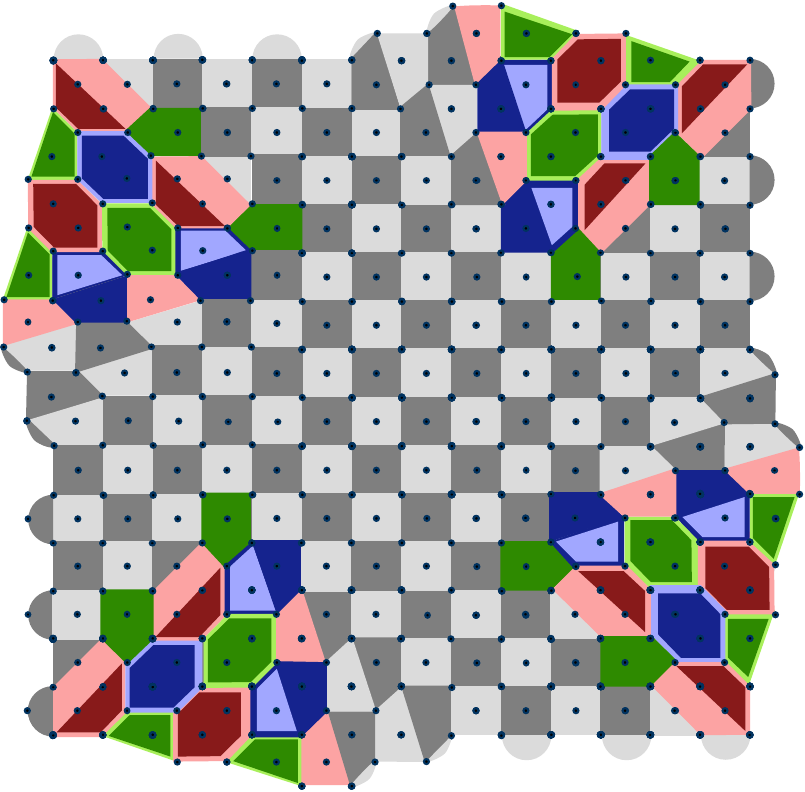}}
    \caption{
    Explicit two-dimensional lattice embedding of the multiplexed injection-and-cultivation layout within a distance-15 logical patch. The outer gray patch represents the full logical patch supporting the final escaped output, while the colored corner regions indicate the four local color-code growth regions used during the early stages of the protocol. Each local site starts from a distance-3 injection region and is subsequently grown to a distance-5 cultivation region. The four corner allocations are symmetry-related embeddings of the same local growth primitive, arranged so that the local regions remain nonoverlapping while leaving a central idle region available during injection and cultivation.
    }
    \label{fig:2DLattice}
\end{figure*}

Fig.~\ref{fig:2DLattice} shows the explicit two-dimensional lattice realization of the multiplexed injection-and-cultivation layout used in this work. The outer patch corresponds to the distance-15 logical patch that ultimately supports the escaped output, while the four colored corner regions correspond to the local color-code growth regions used during the early stages of the protocol. In the proposed scheme, each local site starts from a distance-3 injection region and is subsequently grown to a distance-5 cultivation region. Let \(P^{(15)}\) denote the support of the distance-15 logical patch, and let \(S_i^{(3)}\) and \(S_i^{(5)}\) denote the active supports of the distance-3 and distance-5 local regions for site \(i \in \{1,2,3,4\}\). Then the intended containment relation is
\[
S_i^{(3)} \subset S_i^{(5)} \subset P^{(15)}, \qquad i\in\{1,2,3,4\}.
\]
In the realization used in this paper, the distance-3 and distance-5 local regions require 24 and 53 physical qubits, respectively, whereas the full distance-15 logical patch contains 453 physical qubits. Accordingly, even after embedding four local cultivation regions, a nontrivial idle region remains inside the full patch during the early stages of the protocol.

The key geometric requirement of the multiplexed construction is that the local cultivation regions remain nonoverlapping throughout injection and cultivation. In terms of the active local supports, this condition is expressed as
\[
S_i^{(5)} \cap S_j^{(5)} = \emptyset,
\qquad i \neq j.
\]
The residual idle region is therefore given by
\[
S_{\mathrm{idle}}
=
P^{(15)} \setminus \bigcup_{i=1}^{4} S_i^{(5)}.
\]
This is the geometric basis of the proposed multiplexing scheme: the larger distance-15 patch is not introduced in order to support four independent escaped outputs, but is already required by the standard MSC escape stage; the proposed protocol exploits the fact that only a limited fraction of that eventual footprint is occupied during early-stage growth. As a result, the unused portion of the logical patch can be repurposed to host multiple local growth trajectories in parallel.

The growth of each local color-code region follows the same boundary-aware rule used in standard MSC. Specifically, when a local color code is grown, the newly added data qubits are prepared into Bell pairs, and these Bell pairs are positioned so as to determine the stabilizers whose colors match the boundary being extended~\cite{PhysRevA.93.052332}. Because the Bell-pair preparation acts only on newly added data qubits, the growth step can overlap with the preceding checking step in the original MSC construction~\cite{gidney_magic_2024}. In the present work, this same local growth primitive is reused at all four sites. 

To formalize this reuse, let \(L_{\mathrm{ref}}\) denote a reference local layout. The four corner allocations are obtained through rotation-related embeddings
\[
L_i = R_{\theta_i}(L_{\mathrm{ref}}), 
\qquad
\theta_i \in \{0^\circ, 90^\circ, 180^\circ, 270^\circ\}.
\]
Equivalently, the corresponding local supports satisfy
\[
S_i^{(d)} = R_{\theta_i}\!\left(S_{\mathrm{ref}}^{(d)}\right),
\qquad d \in \{3,5\}.
\]
These symmetry-related embeddings preserve the same local operational role at each site while adapting the local boundary orientation to the corresponding corner of the common distance-15 patch. The purpose of the rotation is therefore not merely aesthetic or geometric: it ensures that the same local color-code growth rule can be applied consistently at all four corner allocations without violating the nonoverlap condition.

This rotation pattern is also tied to the later escape stage. In standard MSC, the cultivated local trajectory is not simply enlarged into a larger color code; instead, it is transferred through a grafted transition into a larger escaped patch. The standard MSC work characterizes this grafted code as a partially folded surface code and emphasizes that grafting is topologically distinct from ordinary lattice surgery~\cite{gidney_magic_2024}. In particular, grafting attaches two boundaries of the color-code region to the larger patch, yields opposite boundary types at the attached corners, and can produce an intermediate code whose distance exceeds that of the small color-code region~\cite{gidney_magic_2024}. For this reason, the local orientation chosen at each corner must remain compatible with a subsequent grafted escape path. In the proposed multiplexed protocol, only one surviving local trajectory is eventually selected for continuation, but every corner allocation is constructed so that, if selected, it can be connected to the same escape framework in a geometrically consistent manner.

From this viewpoint, the proposed four-site multiplexed layout should be understood as a rotated reuse of a single local cultivation primitive within the idle portion of the distance-15 logical patch. The local sites are not four distinct constructions; they are four symmetry-related embeddings of the same injection-and-cultivation primitive, arranged so as to satisfy containment, nonoverlap, and later escape compatibility simultaneously.

\section{Implementation details for candidate selection before escape}
\label{app:selective_escape_detail}

This appendix provides a more detailed implementation of the selective escape procedure used in the proposed multiplexed MSC protocol. The goal is to formalize how detector outcomes are converted into site-wise success or failure decisions, how the surviving local sites are assembled into a candidate set, and how one selected trajectory is forwarded to the standard MSC escape framework. The construction follows the overall stage-dependent philosophy of standard MSC: early-stage detector events are handled by postselection, whereas the final escape stage is treated using a decoder-aware acceptance rule.

\subsection{Detector-based success criterion and candidate formation}

Let the four local cultivation sites be indexed by \(i \in \{1,2,3,4\}\).
For each site \(i\), define two binary survival indicators:
\[
\chi_i^{(\mathrm{inj})},\ \chi_i^{(\mathrm{cult})} \in \{0,1\}.
\]
Here, \(\chi_i^{(\mathrm{inj})}=1\) means that site \(i\) survives injection without triggering an error-detection event, and \(\chi_i^{(\mathrm{cult})}=1\) means that site \(i\) survives cultivation without triggering an error-detection event. Otherwise, the corresponding indicator is \(0\).

% ~\textcolor{red}{
% Let the four local cultivation sites be indexed by \(i \in \{1,2,3,4\}\). For each site, define binary survival indicators
% \[
% \chi_i^{(\mathrm{inj})}, \chi_i^{(\mathrm{cult})} \in \{0,1\},
% \]
% where
% \[
% \chi_i^{(\mathrm{inj})} =
% \begin{cases}
% 1, & \text{if no error-detection event is triggered at site } i \text{ during injection},\\
% 0, & \text{otherwise},
% \end{cases}
% \]
% and
% \[
% \chi_i^{(\mathrm{cult})} =
% \begin{cases}
% 1, & \text{if no error-detection event is triggered at site } i \text{ during cultivation},\\
% 0, & \text{otherwise}.
% \end{cases}
% \]}
The overall early-stage survival indicator for site \(i\) is then
\[
\chi_i = \chi_i^{(\mathrm{inj})}\chi_i^{(\mathrm{cult})}.
\]

Using these indicators, we define the candidate set
\[
\mathcal{C}
=
\left\{
i \in \{1,2,3,4\}
\;\middle|\;
\chi_i = 1
\right\}.
\]
That is, a site belongs to \(\mathcal{C}\) if and only if it completes both injection and cultivation without triggering an error-detection event. This rule is the multiplexed analogue of the full-postselection policy used in the early stages of standard MSC, where any detector firing during injection or cultivation causes the corresponding trial to be discarded. In the present work, the same logic is localized to each site: instead of discarding the entire shot immediately, only the site that triggers the detector event is removed from further advancement, while the remaining sites continue through the early-stage protocol. This retains the simplicity of detector-based rejection while allowing multiple local trials to coexist within one shot. 

This adaptation is also motivated by the same feedback consideration emphasized in the standard MSC. In the original protocol, full postselection during injection and cultivation avoids the need for complex just-in-time decoding to correct detected errors before they corrupt the cultivated state. In the multiplexed setting, the same advantage is preserved site-wise: once a detector fires at a given site, that site is marked as failed and removed from further growth, so no urgent correction is needed for that site. The remaining sites continue unaffected.

\subsection{Selection rule and idle treatment}

Once the candidate set \(\mathcal{C}\) has been formed, the protocol distinguishes two cases. If
\[
\mathcal{C} = \emptyset,
\]
then no local trajectory has survived the early stages, and the shot is discarded. A new shot must then be started from the beginning. If instead
\[
|\mathcal{C}| \ge 1,
\]
then one candidate is selected for continuation into the escape stage. We denote the selected site by
\[
c^\star \in \mathcal{C}.
\]
The selection map
\[
\Pi: 2^{\{1,2,3,4\}} \setminus \{\emptyset\} \to \{1,2,3,4\}
\]
is assumed to be a predetermined rule satisfying
\[
c^\star = \Pi(\mathcal{C}), \qquad c^\star \in \mathcal{C}.
\]
In the simplest implementation, \(\Pi\) may be chosen as a fixed deterministic priority rule, e.g., selecting the lowest-index surviving site. More generally, any predetermined rule is acceptable so long as exactly one surviving candidate is selected.

After selection, all sites \(i \neq c^\star\) are marked inactive for the remainder of the shot. This includes both failed sites \((\chi_i=0)\) and surviving-but-unselected sites \((i\in\mathcal{C}\setminus\{c^\star\})\). Formally, the continuation state of each site may be written as
\[
\eta_i =
\begin{cases}
1, & i = c^\star,\\
0, & i \neq c^\star,
\end{cases}
\]
where \(\eta_i=1\) indicates that site \(i\) continues into escape and \(\eta_i=0\) indicates that the site remains idle or inactive. In this way, the selective escape procedure maps multiple early-stage local trials onto a single continuation path for final logical-state generation.

\subsection{Connection to the standard MSC escape framework}

Once the selected site \(c^\star\) has been determined, the protocol returns to the standard MSC escape framework. The selected local trajectory is forwarded to the same grafted transition and subsequent escape path described for the single-site protocol in the main text. Thus, multiplexing is confined to the early injection-and-cultivation stages, whereas the escape stage remains a single-path continuation acting on one selected local trajectory.

In standard MSC, the decoding treatment of the escape stage differs qualitatively from the early stages. Injection and cultivation use full postselection, but escape is too large to postselect every possible detection event. Instead, the decoding problem is transformed from a color-code problem into a matching problem by postselecting detection events on selected stabilizers in the color-code region, specifically red X-basis stabilizers and green Z-basis stabilizers in the original construction. This guarantees that the number of unpostselected detection events produced by a single \(X\)- or \(Z\)-type data error remains at most two, allowing the escape stage to be exposed to a matching decoder. The standard MSC construction further introduces doublet nodes in order to represent otherwise unmatchable excitations inside the color-code region~\cite{gidney_magic_2024}.

The proposed multiplexed protocol does not modify this escape-stage decoding principle. Instead, it applies it only to the selected site \(c^\star\). In other words, selective escape determines which local trajectory is allowed to enter the decoder-aware part of the protocol, while the subsequent acceptance or rejection of that trajectory follows the same escape-stage logic as in standard MSC. If desired, the final keep/discard decision for the selected site may be expressed as a decoder-dependent indicator
\[
\kappa(c^\star) \in \{0,1\},
\]
where \(\kappa(c^\star)=1\) indicates that the selected trajectory is accepted as a valid output after escape, and \(\kappa(c^\star)=0\) indicates that it is rejected.

In the original MSC construction, the decoder confidence used for this final acceptance decision is quantified by a complementary gap. This suggests a natural interpretation in the multiplexed setting: site-wise selection determines which surviving local trajectory enters escape, while decoder-based confidence determines whether the selected trajectory is ultimately kept as the final logical magic-state output. Thus, the selective escape procedure and the escape-stage decoder serve complementary roles: the former chooses one surviving local trajectory, and the latter decides whether that trajectory is sufficiently reliable to keep.

\section{\label{sec:Gap-based acceptance and distance-dependent logical-error behavior}Gap-based acceptance and distance-dependent logical-error behavior}

This appendix provides a detailed interpretation of the logical-error behavior observed in Fig.~\ref{fig:E2ECost}. As discussed in the main text, the primary effect of the proposed multiplexed MSC protocol is to reduce discards during the early injection and cultivation stages. However, a final kept shot is obtained only after the selected candidate is forwarded to the escape stage and satisfies the decoder gap criterion. Therefore, the end-to-end behavior in Fig.~\ref{fig:E2ECost} reflects both the reduction of early-stage rejection and the gap-based acceptance applied after escape.

After the escape stage, the decoder compares the weights of the two logical hypotheses associated with a given syndrome. We denote the gap between these two hypotheses by
\begin{equation}
\Delta = |w_0 - w_1|,
\end{equation}
where \(w_0\) and \(w_1\) are the decoder weights for the two logical hypotheses. A larger value of \(\Delta\) indicates that the decoder more clearly favors one hypothesis over the other, whereas a smaller value indicates that the two hypotheses are closer in weight.

\begin{figure*}[!t]
    \centering{\includegraphics[width=17.5cm]{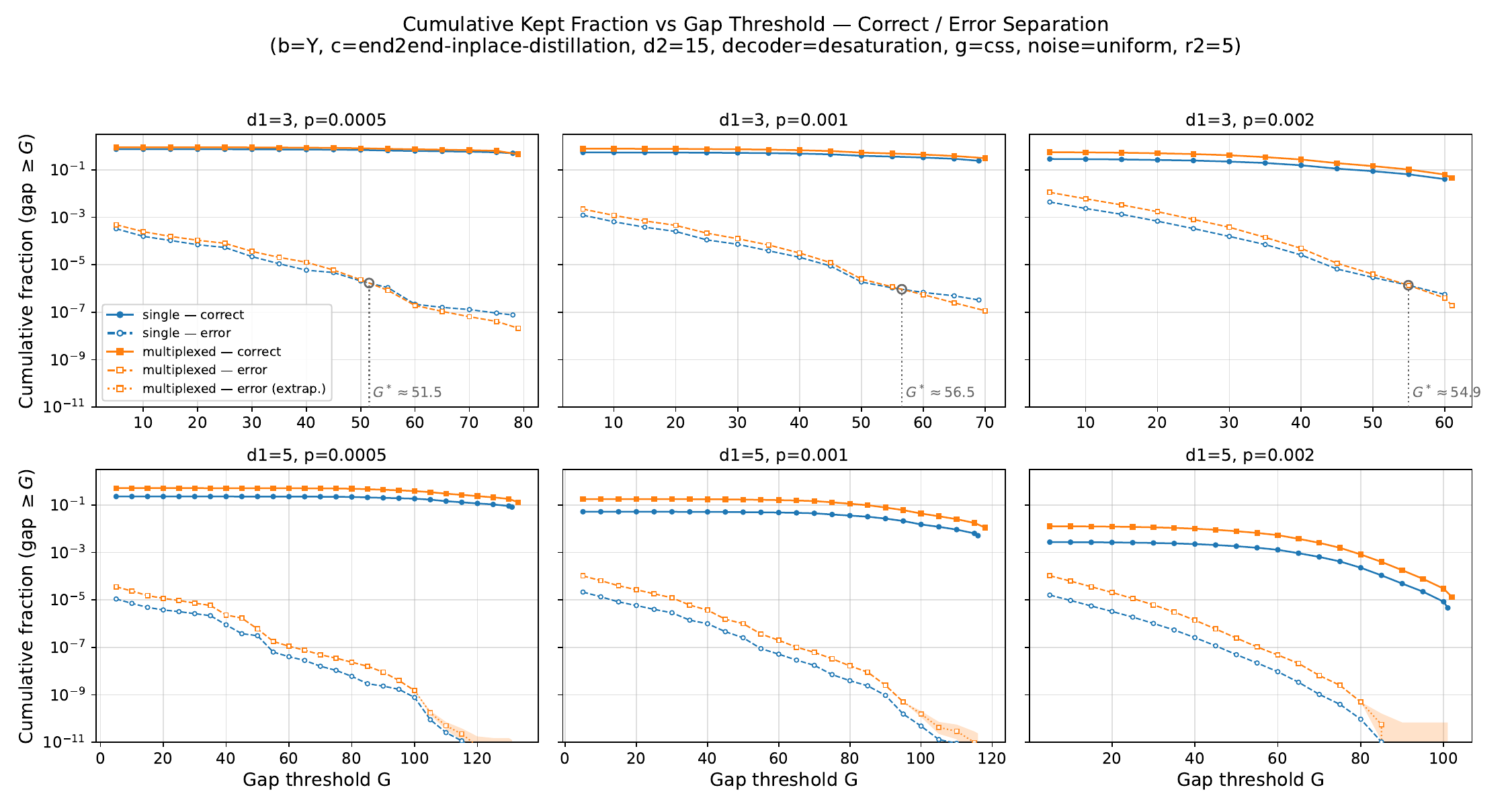}}
    \caption{
    Cumulative fractions of correct and erroneous shots remaining after applying the gap threshold \(G\). Solid lines denote correct shots, dashed lines denote erroneous shots, blue indicates the single-site MSC baseline, and orange indicates the multiplexed MSC protocol. For \(d_1=3\), the multiplexed protocol achieves a lower logical error rate at larger \(G\), whereas for \(d_1=5\), it mainly reduces the expected number of attempts while maintaining a comparable logical error rate. The shaded regions indicate extrapolated estimates.
    }
    \label{fig:GapCumulativeDistribution}
\end{figure*}

For a given gap threshold \(G\), only shots satisfying \(\Delta \ge G\) are counted as final kept shots. Let \(N_C(G)\) and \(N_E(G)\) denote the numbers of correct and erroneous shots, respectively, that remain after applying this criterion. The logical error rate after gap-based acceptance is then
\begin{equation}
p_L(G)
=
\frac{N_E(G)}
{N_C(G)+N_E(G)},
\label{eq:gap_ler}
\end{equation}
and the expected number of attempts per kept shot is
\begin{equation}
A(G)
=
\frac{N_{\mathrm{shots}}}
{N_C(G)+N_E(G)} .
\label{eq:gap_attempts}
\end{equation}
Thus, the logical error rate is not determined by the absolute number of erroneous shots alone, but by the fraction of erroneous shots among all shots remaining after the gap criterion. Increasing \(G\) reduces the number of final kept shots and therefore increases \(A(G)\). At the same time, shots for which the decoder has a smaller separation between the two logical hypotheses are removed, so \(p_L(G)\) generally decreases.

It is important to distinguish the early-stage candidate selection rule from the escape-stage acceptance rule. In the proposed multiplexed protocol, the candidate set is formed only from detector outcomes during injection and cultivation. If the early-stage pass indicator for site \(i\) is written as
\begin{equation}
\chi_i
=
\chi_i^{(\mathrm{inj})}
\chi_i^{(\mathrm{cult})},
\end{equation}
then the candidate set is
\begin{equation}
\mathcal{C}
=
\{i \in \{1,2,3,4\} \mid \chi_i = 1\}.
\end{equation}
At this stage, the escape operation has not yet been performed, so the decoder gap \(\Delta\) is not available. The selected site \(c^\star\) is therefore chosen from candidates that passed injection and cultivation without using gap information. The gap is computed only after the selected candidate undergoes escape and is decoded. Consequently, the logical-error behavior in Fig.~\ref{fig:E2ECost} should be interpreted as the result of the post-escape gap criterion, rather than as a consequence of gap-aware candidate selection.

Figure~\ref{fig:GapCumulativeDistribution} separates the two quantities that determine the logical error rate. Solid curves show the fraction of correct shots satisfying \(\Delta \ge G\), while dashed curves show the fraction of erroneous shots satisfying the same condition. Blue curves correspond to the single-site MSC baseline, and orange curves correspond to the multiplexed protocol. The plotted quantities are normalized by the total number of attempts. As \(G\) increases, both correct and erroneous shots decrease because the acceptance condition becomes stricter. The final logical error rate is determined by the relative composition of the correct and erroneous shots that remain after this criterion.

This behavior depends on the local cultivation distance. In the distance-3 case, a crossing appears between the single-site MSC and multiplexed MSC results as the gap threshold is increased. This crossing corresponds to the threshold at which the logical error rate of the multiplexed protocol becomes equal to that of the single-site baseline. Beyond this threshold, the multiplexed protocol exhibits a lower logical error rate than the single-site MSC baseline. This occurs because, after applying the gap criterion, the fraction of erroneous shots among the remaining shots becomes smaller for the multiplexed protocol than for the single-site baseline. In other words, as the gap threshold is increased in the distance-3 case, shots containing errors are effectively removed, while the multiplexed protocol retains enough correct shots due to reduced early-stage discard. Therefore, for distance 3, early-stage discard reduction and escape-stage gap-based acceptance act together to produce a regime beyond the crossing where both the logical error rate and the expected number of attempts are reduced. 

The distance-3 results further show that the crossing occurs at a noticeably smaller gap threshold for the lowest physical error rate considered. For \(p=0.0005\), the crossing is observed at \(G^\star \approx 51.5\), compared with \(G^\star \approx 56.5\) for \(p=0.001\) and \(G^\star \approx 54.9\) for \(p=0.002\). This indicates that, in the lower-noise case, a weaker gap criterion is sufficient for the multiplexed protocol to enter the lower-logical-error regime while retaining the reduced retry cost from early-stage multiplexing.

In contrast, the distance-5 case shows a different behavior. As shown in the distance-5 panels of Fig.~\ref{fig:GapCumulativeDistribution}, the multiplexed protocol retains more correct shots than the single-site baseline after applying the gap criterion, thereby reducing the expected number of attempts required to obtain a kept output. However, erroneous shots also remain, so the fraction of erroneous shots among all shots remaining after the gap criterion does not become sufficiently smaller than that of the single-site baseline. As a result, in the distance-5 case, the logical error rate does not show the same pronounced post-crossing improvement observed in the distance-3 case. Instead, the benefit of multiplexing appears more clearly as a reduction in expected attempts while maintaining comparable logical-error behavior. In other words, for distance 5, multiplexing lowers the retry cost by forwarding more candidates to the escape stage, while the logical error rate of the final kept shots remains at a level similar to that of the single-site baseline.

This comparison highlights an important point in interpreting the effect of multiplexing. The most consistent effect of multiplexing is to reduce the probability that the entire shot is discarded during the early injection and cultivation stages, thereby reducing the expected number of attempts per kept shot. In contrast, the logical error rate is determined by the correct/error composition of the shots that remain after applying the post-escape gap criterion. Therefore, logical-error improvement is not guaranteed by multiplexing alone, but depends on the local cultivation distance, the physical error rate, and the chosen gap threshold.

The shaded regions in Fig.~\ref{fig:GapCumulativeDistribution} indicate extrapolated estimates in the rare-event tail. At large \(G\), erroneous shots are observed only rarely. Therefore, the tail region should be interpreted as a supplementary estimate of how erroneous shots decrease under stronger gap filtering, rather than as a direct count at every threshold.

\clearpage

\bibliography{manuscript}% Produces the bibliography via BibTeX.

\end{document}